\documentstyle[prl,aps,floats,psfig]{revtex} 
\begin{document}
\twocolumn[
\hsize\textwidth\columnwidth\hsize\csname@twocolumnfalse\endcsname


\title
{Violation of Quasineutrality in Semiconductor Transport: The Dember Effect}

\author{Maja Kr\v cmar and Wayne M. Saslow}
\address{Department of Physics, Texas A\&M University, College Station,
Texas 77840-4242}


\maketitle

\begin{abstract}Exact solution of the linearized equations for steady-state
transport in semiconductors yields two modes that vary exponentially in space,
one involving screening (without entropy production) and one involving diffusion
and recombination (with entropy production).  Neither mode is quasineutral.  For
constant surface photoexcitation with generation of electrons and holes, the
steady-state response is a linear combination of these modes, subject to global
electroneutrality. The resultant charge separation produces a voltage difference
across the sample (the Dember effect).


\end{abstract}
]


The Dember effect is the voltage difference that develops across the bulk of a
material that steadily absorbs light at one surface.\cite{dember}  It has been
observed in insulators,\cite{ber1,ber2} high resistivity
semiconductors,\cite{bad,berg} and $C_{60}$ thin films.\cite{dipan} An extensive
discussion of photovoltages, both at surfaces and in the bulk, is given in
Ref.\onlinecite{kronikshap}.

An intuitive physical picture of the Dember effect is as follows.\cite{chaz} 
Incident light produces equal numbers of electrons and holes at the surface. The
higher mobility electrons travel further than the lower mobility holes.  The
resultant charge separation amounts to a dipole layer.  As a consequence the
illuminated surface is at the higher voltage, in agreement with experiment.
However, conventional theories of the Dember effect,\cite{kronikshap,chaz}
which assume quasineutrality (local electroneutrality), yield no dipole layer;
moreover, they yield a non-zero net charge.

The present work obtains exact solutions for the surface modes of the linearized
macroscopic transport equations.  It differs from previous analytic work in
that: (1) It explicitly satisfies the principles of irreversible thermodynamics,
which constrain the form for the charge-carrier fluxes and recombination rates
by the condition that the rate of entropy production be non-negative; (2) In
solving the transport equations, it employs both electron and hole densities $n$
and $p$ as variables, rather than assuming deviations from equilibrium satisfy
quasineutrality ($\delta n\approx \delta p$); (3) Using the resultant two
steady-state surface modes, it explicitly enforces overall electroneutrality.

One surface mode corresponds to ordinary screening by a multi-charge-carrier
system (i.e. Debye-H\"uckel screening), without fluxes or entropy production:
the {\it screening mode.} The other surface mode has both electron and hole flux
as well as entropy production: the {\it diffusion-recombination mode.} The
characteristic length $l$ of the screening mode typically is shorter than the
characteristic length $L$ for the diffusion-recombination mode.  Neither mode is
locally electroneutral.  Applied to the Dember effect, the light generates a
diffusion-recombination mode with a relatively extended negative charge and a
screening mode with a relatively compact positive charge, in agreement with the
physical picture given above.

{\bf Irreversible thermodynamics.} Consider a uniform semiconductor with an
ideal surface having no extrinsic surface states and no charged intrinsic
states. The recombination rate $r$ is the same for both electrons and holes. 
With $u$ the energy density, $T$ the temperature, $s$ the entropy density, and
$\tilde\mu_e$ and $\tilde\mu_h$ the electron and hole electrochemical
potentials, the fundamental thermodynamic differential for this system is
\begin{equation}
du=Tds+\tilde\mu_e dn+\tilde\mu_h dp.
\label{1}
\end{equation}
With $\mu_e$ and $\mu_h$ the chemical potentials (to be distinguished from the
diffusivities $\mu_n$ and $\mu_p$),
\begin{equation}
\tilde\mu_e\equiv\mu_e-e\phi, \qquad \tilde\mu_h\equiv\mu_h+e\phi.  
\label{2}
\end{equation}
Here the electrical potential $\phi$ satisfies Poisson's equation
\begin{equation}
\nabla^2\phi=-{\rho\over\epsilon_0\epsilon}
=-{e\over\epsilon_0\epsilon}(p-n+N_d^+ - N_a^-),
\label{3}
\end{equation}
where $\epsilon$ is the semiconductor dielectric constant, $\epsilon_0$ is the
permittivity of free space, and the charge density $\rho=e(p-n+N_d^+ - N_a^-)$.
$N_d^+$ and $N_a^-$ are the respective densities of ionized donors and
acceptors, and $\tilde\mu_e$ and $\tilde\mu_h$ are often called quasi-Fermi
energies.

The conservation laws for this system are
\begin{equation}
\partial_t u+\partial_i j_i^u=0, \quad
\partial_t s+\partial_i j_i^s={{\cal P}\over T}\ge 0, 
\label{4}
\end{equation}
\begin{equation}
\partial_t n+\partial_i j_i^n=r, \quad \partial_t p+\partial_i j_i^p=r. 
\label{5}
\end{equation}
Here $j_i^u$ is the energy flux density, $j_i^s$ is the entropy flux
density, ${\cal P}$ is the local rate of heat production (${\cal P}/T$ is local
rate of entropy production), $j_i^n$ is the electron number flux density, and
$j_i^p$ is the hole number flux density.  The fluxes, $r$, and $\cal P$ are to
be determined.

The time-derivatives of (\ref{1}) and (\ref{4})--(\ref{5}) lead to 
\begin{eqnarray}
0\le {\cal P}&=&-\partial_i j_i^u +T\partial_i j_i^s
-\tilde\mu_e(r-\partial_i j_i^n)-\tilde\mu_h(r-\partial_i j_i^p)
\nonumber\\
&=&-\partial_i(j_i^u - Tj_i^s - \tilde\mu_e j_i^n - \tilde\mu_h
j_i^p)\nonumber\\
& &-j_i^s\partial_i T - j_i^n\partial_i\tilde\mu_e - j_i^p\partial_i\tilde\mu_h
-r(\tilde\mu_e+\tilde\mu_h). 
\label{6}
\end{eqnarray}

Expressing ${\cal P}$ as a non-negative quadratic form requires that 
$j_i^u=Tj_i^s+\tilde\mu_e j_i^n+\tilde\mu_h j_i^p$; that 
\begin{equation}
j_i^s =-\alpha_T\partial_iT=-{\kappa\over T}\partial_iT, 
\quad 
\label{7}
\end{equation}
where $\kappa\ge 0$ is the thermal conductivity; that  
\begin{eqnarray}
j_i^n=-\alpha_{nn}\partial_i\tilde\mu_e-\alpha_{np}\partial_i\tilde\mu_h,\nonumber\\
j_i^p=-\alpha_{pn}\partial_i\tilde\mu_e-\alpha_{pp}\partial_i\tilde\mu_h, 
\label{8}
\end{eqnarray}
where $\alpha_{nn}\ge 0, \alpha_{pp}\ge 0$, and 
$\alpha_{nn}\alpha_{pp}\ge \alpha_{np}\alpha_{pn}$ (the $\alpha$'s 
are related to the diffusivities, and $\alpha_{np}=\alpha_{pn}$ by the Onsager 
symmetry principle\cite{Onsager-LL}); and that 
\begin{equation}
r=-\lambda(\tilde\mu_e+\tilde\mu_h)=-\lambda(\mu_e+\mu_h). 
\label{9}
\end{equation}
where $\lambda\ge 0$ is related to the electron and hole lifetimes $\tau_n$ and
$\tau_p$. In equilibrium $r=0$, so (\ref{9}) implies the well-known results
that $\tilde\mu_e=-\tilde\mu_h$ and $\mu_e=-\mu_h$ in equilibrium.  We
now express $r$, $j_i^n$, and $j_i^p$ in more conventional form.  

Linearizing (\ref{9}) about equilibrium with $\delta
\mu_e=(\partial\mu_e/\partial n)\delta n$ and $\delta
\mu_h=(\partial\mu_h/\partial p)\delta p$ yields
\begin{equation}
r=-{\delta n\over\tau_n} -{\delta p\over\tau_p}, 
\quad \tau_n^{-1}\equiv \lambda{\partial\mu_e\over\partial n}, 
\quad \tau_p^{-1}\equiv \lambda{\partial\mu_h\over\partial p}.
\label{10}
\end{equation}
For a dilute semiconductor, as considered here, 
$(\partial\mu_e/\partial n)\approx k_B T/n_0$ and 
$(\partial\mu_h/\partial p)\approx k_B T/p_0$, so
$\tau_n=(n_0/k_B T\lambda)$ and $\tau_p=(p_0/k_B T\lambda)$.  

Now set $\alpha_{np}=\alpha_{pn}\approx 0$,  
$\alpha_{nn}=D_n(\partial n/\partial \mu_e)$, and 
$\alpha_{pp}=D_p(\partial p/\partial \mu_h)$, where $D_n$
and $D_p$ are the electron and hole diffusivities.  Next, define 
the electron and hole mobilities $\mu_n$ and $\mu_p$, which satisfy 
the Einstein relations: 
\begin{equation}
\mu_n=
{eD_n\over n}{\partial n\over\partial \mu_e}\approx {eD_n\over k_B T}, 
\quad \mu_p=
{eD_p\over p}{\partial p\over\partial \mu_h}\approx {eD_p\over k_B T}, 
\label{11}
\end{equation}
Then in one-dimension $j_i^n$ and $j_i^p$ become\cite{ash,sze}  
\begin{eqnarray}
j^n_x=-D_n\partial_x n+\mu_n n\partial_x\phi, \nonumber\\
j^p_x=-D_p\partial_x p-\mu_p p\partial_x\phi. 
\label{12}
\end{eqnarray}

Note that Poisson's equation linearizes to 
\begin{eqnarray}
\nabla^2\delta\phi&=&-{e\over\epsilon_0\epsilon}(\delta p-\delta n)
={e^2\over\epsilon_0\epsilon}({\partial n\over \partial\mu_e}+
{\partial p\over \partial\mu_h})\delta\phi\nonumber\\
& &-{e\over\epsilon_0\epsilon}({\partial p\over \partial\mu_h}\delta\tilde\mu_h
-{\partial n\over \partial\mu_e}\delta\tilde\mu_e).
\label{13}
\end{eqnarray}


{\bf Static surface solutions in one-dimension.} One steady-state solution that
automatically satisfies (\ref{5}), (\ref{8}), and (\ref{9}) has
$\delta\tilde\mu_e=\delta\tilde\mu_h=0$ and $j_i^n=j_i^p=0$.  Thus the system
is in local equilibrium, with no entropy production.  This corresponds to
ordinary screening, and thus we call it the {\it screening mode,} with subscript
$S$.  From (\ref{13}), its potential $\delta\phi_S$ satisfies
\begin{equation}
\nabla^2\delta\phi_S
={e^2\over\epsilon_0\epsilon}({\partial p\over \partial\mu_h}+
{\partial n\over \partial\mu_e})\delta\phi_S=q_S^2\delta\phi_S, 
\label{14}
\end{equation}
where 
\begin{equation}
q_S^2=q_{Sn}^2+q_{Sh}^2, 
\quad q_{Sn}^2\equiv{e^2\over\epsilon_0\epsilon}{\partial n\over \partial\mu_e}, 
\quad q_{Sp}^2\equiv{e^2\over\epsilon_0\epsilon}{\partial p\over \partial\mu_h}.
\label{15}
\end{equation}
For the present system, $q_{Sn}^2=(e^2n_0/\epsilon_0\epsilon k_B T)$,
$q_{Sp}^2=(e^2p_0/\epsilon_0\epsilon k_B T)$, so $q_S^2=(e^2/\epsilon_0\epsilon
k_B T)(n_0+p_0)$.  The solution to (\ref{14}) that goes to zero as
$x\rightarrow\infty$ is
\begin{equation}
\delta\phi_S=A_S\exp{(-q_S x)}. 
\label{16}
\end{equation}
$l\equiv q_S^{-1}$ is the {\it screening length.} From (\ref{16}) and
(\ref{13}), the screening mode has charge density
\begin{equation}
\delta\rho_S=-\epsilon_0\epsilon\nabla^2\phi_S
=-(\epsilon_0\epsilon) q_S^2 A_S\exp{(-q_S x)}.
\label{17}
\end{equation}
Here $\delta n_S=(\partial n/\partial\mu_e)\delta\mu_e
=(en_0/k_B T)\delta\phi_S$ and 
$\delta p_S=(\partial n/\partial\mu_h)\delta\mu_h
=-(en_0/k_B T)\delta\phi_S$, so $\delta n_S/\delta p_S=-n_0/p_0$. 


The second steady-state solution to (\ref{5}), (\ref{8}), (\ref{9}) and
(\ref{13}) has $\delta\tilde\mu_e$ and $\delta\tilde\mu_h$ non-zero, so the
fluxes $j_i^n$ and $j_i^p$ are non-zero.  With $\alpha_{np}=\alpha_{pn}=0$,
combining (\ref{5}), (\ref{8}) and (\ref{9}) yields
\begin{eqnarray}
-\alpha_{nn}\partial_x^2\delta\tilde\mu_e
=-\lambda(\delta\tilde\mu_e+\delta\tilde\mu_h), \nonumber\\
-\alpha_{pp}\partial_x^2\delta\tilde\mu_h
=-\lambda(\delta\tilde\mu_e+\delta\tilde\mu_h).
\label{18}
\end{eqnarray}
Let us denote properties of this {\it diffusion-recombination mode} by the
subscript $D$.  Now set
\begin{equation}
q_{Dn}^2\equiv {\lambda\over\alpha_{nn}}={1\over D_n\tau_n}, \quad
q_{Dp}^2\equiv {\lambda\over\alpha_{pp}}={1\over D_p\tau_p}.   
\label{19}
\end{equation}
Then, with
\begin{equation}
\delta\tilde\mu_e=A_{Dn}\exp{(-q_D x)}, \quad \delta\tilde\mu_h=A_{Dp}\exp{(-q_D x)},
\label{20}
\end{equation}
(\ref{18}) yields  
\begin{equation}
q_D^2=q_{Dn}^2+q_{Dp}^2  
\label{21}
\end{equation}
and $A_{Dp}/q_{Dp}^2=A_{Dn}/q_{Dn}^2$. 

The corresponding electrostatic potential $\delta\phi_D$ takes the form 
\begin{equation}
\delta\phi_D=A_D\exp{(-q_D x)},
\label{22}
\end{equation}
where 
\begin{equation}
{A_{Dp}\over q_{Dp}^2}={A_{Dn}\over q_{Dn}^2}
={q_S^2-q_D^2\over q_{Sp}^2q_{Dp}^2-q_{Sn}^2q_{Dn}^2}eA_D.  
\label{23}
\end{equation}
$L\equiv q_D^{-1}$ is the {\it diffusion-recombination length,} normally
called the diffusion length.  By (22) and (3), the diffusion-recombination mode
possesses charge density
\begin{equation}
\delta\rho_D=-(\epsilon_0\epsilon) q_D^2 A_D\exp{(-q_D x)}.
\label{24}
\end{equation}
Moreover,  
$\delta n_D/\delta p_D=(n_0/p_0)(q_{Sp}^2-q_{Dn}^2)/(q_{Sn}^2-q_{Dp}^2)$.  This 
is independent of $n_0$ for fixed $n_0p_0=n_i^2$. 


{\bf Surface mode amplitudes.}  Let light incident at the surface $x=0$ 
produce equal electron and hole fluxes $G$, so $j_i^n=j_i^p=G$ at $x=0$.  On
setting $\alpha_{np}=\alpha_{pn}=0$, (\ref{8}) and (\ref{20}) yield the
diffusion-recombination mode amplitudes
\begin{equation}
A_{Dn}={G\over q_D\alpha_{nn}}, \quad A_{Dp}={G\over q_D\alpha_{pp}}. 
\label{25}
\end{equation}

Overall electroneutrality, or
$0=\int_0^{\infty}dx(\delta\rho_S(x)+\delta\rho_D(x))$, implies that 
\begin{equation}
A_S=-{q_D\over q_S}A_D.
\label{26}
\end{equation}


From (\ref{17}) and (\ref{26}), and from (\ref{23}-\ref{25}), the total charge
density is
\begin{equation}
\delta\rho(x)={Ge^2\over k_BT}\ {(\mu_n-\mu_p)\over \mu_n\mu_p}\ 
{q_S e^{-q_S x}-q_D e^{-q_D x}\over q_S^2-q_D^2}, 
\label{27}
\end{equation}
where the mobilities have been employed.  Typically $\mu_n>\mu_p$, so
$\rho(x=0)$ is positive, as expected if the higher mobility charge-carrier
preferentially leaves the vicinity of the surface.  Note that $\delta\rho(x)$
changes in sign as $x$ increases, as needed to produce a dipole layer.

From (\ref{17}) and (\ref{24}), the total electrostatic potential is 
\begin{equation}
\delta\phi(x)={Ge^2\over \epsilon_0\epsilon k_BT}\ 
{(\mu_n-\mu_p)\over \mu_n\mu_p}\ 
{q_D^{-1}e^{-q_D x}-q_S^{-1}e^{-q_S x}\over q_S^2-q_D^2}.
\label{28}
\end{equation}

To compare with experiment, (\ref{28}) gives the Dember voltage 
$\phi_{Dem}\equiv\delta\phi(x=0)$ as 
\begin{equation}
\phi_{Dem}={Ge^2\over \epsilon_0\epsilon k_BT}
\ {(\mu_n-\mu_p)\over \mu_n\mu_p}\ 
{1\over q_S q_D}{1\over q_S+q_D}.
\label{29}
\end{equation}

{\bf Discussion.} Eq.(29) makes numerous predictions. First, since typically
$\mu_n>\mu_p$, $\phi_{Dem}$ is positive, as expected for a dipole layer with
positive charge closer to the surface. Further, since typically $q_D\ll q_S$,
the term $q_S q_D(q_S+q_D)\approx q_S^2 q_D$.  The dependences of $q_S$ on $n$
and $\epsilon$, and of $q_D$ on $\mu$ and $\tau$, show that $\phi_{Dem}$ varies
inversely with carrier density $n$, as the square root of the characteristic
recombination time $\tau$, as the inverse square root of the characteristic
mobility $\mu$, and is independent of $\epsilon$.

The dependence on $n$ explains why $\phi_{Dem}$ is observed in materials with
relatively low carrier densities.  The dependence on $\tau$ indicates that
$\phi_{Dem}$ is larger for longer recombination times; the dependence on $\mu$
indicates that $\phi_{Dem}$ is larger for lower mobilities.  

\begin{figure}
\centerline{\psfig{file=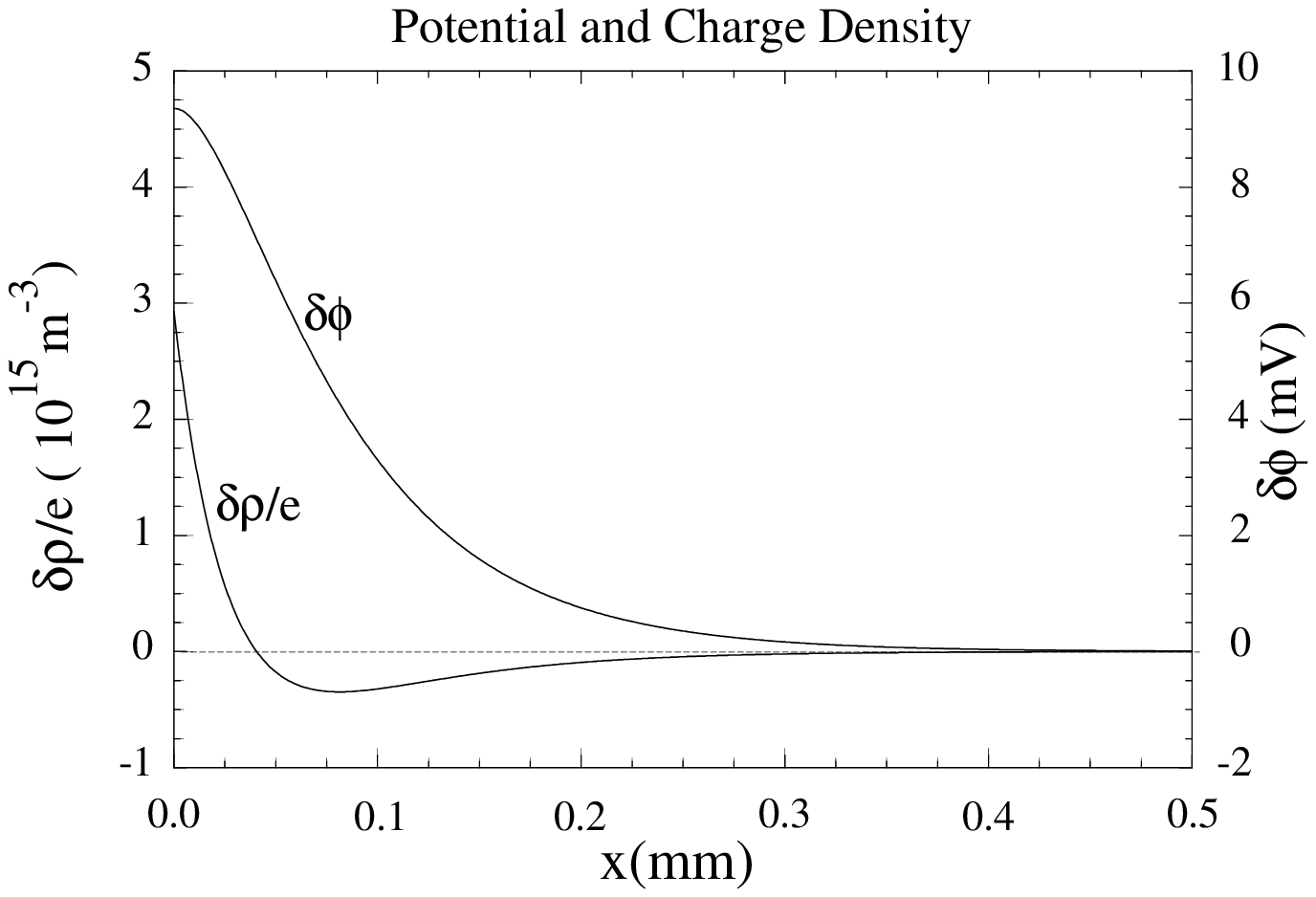,width=1\linewidth}}
\caption{Photo-induced electrical potential $\delta\phi$ from (28) and charge 
density $\delta\rho$ from (27).} 
\label{fig:1}
\end{figure}

\begin{figure}
\centerline{\psfig{file=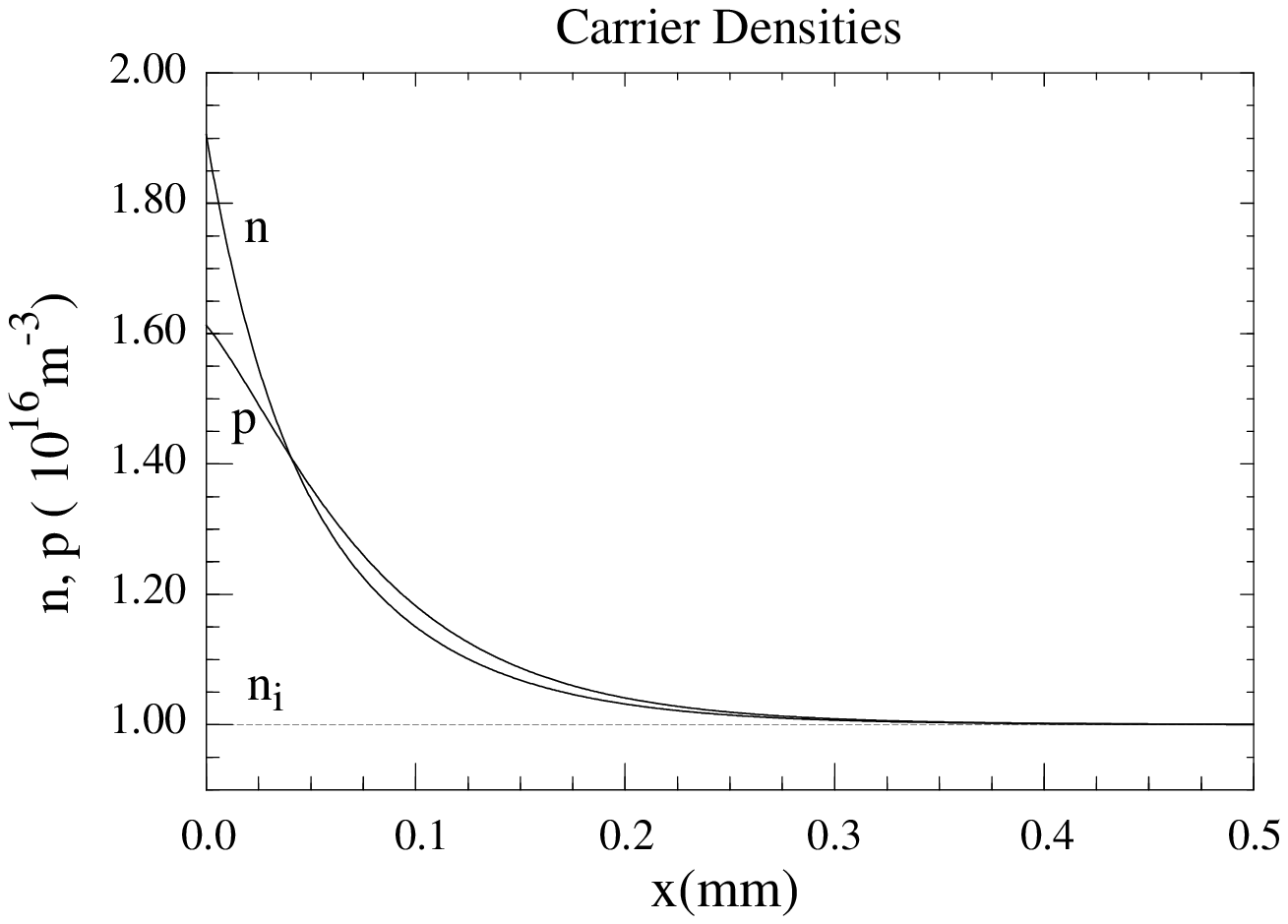,width=1\linewidth}}
\caption{Electron and hole number densities $n$ and $p$ corresponding to Figure 1.} 
\label{fig:2}
\end{figure}

Figures 1-4 are computed for an intrinsic semiconductor with equilibrium carrier
number density $n_0=p_0=n_i=10^{16}$~m$^{-3}$, recombination time
$\tau=10^{-5}$~s, electron mobility $\mu_n=1000$~cm$^{2}$/(V-s), hole mobility
$\mu_p=200$~cm$^{2}$/(V-s), room temperature $k_B T=0.0253$~eV, and dielectric
constant $\epsilon=10$.  These values give $l=26.4~\mu$m and $L=64.8~\mu$m. From
(28), the ratio of the contributions of the diffusion mode and the screening
mode to the Dember voltages is $q_D^{-1}/q_S^{-1}=L/l$. Taking
$G=10^{17}$~m$^{-2}$s$^{-1}$ gives $\phi_{Dem}=9.22$~mV, of which the diffusion
mode is responsible for 15.56~mV and the screening mode $-6.34$~mV.  The theory,
which is valid for $\delta\tilde\mu_e$ and $\delta\tilde\mu_h$ small relative to
$k_B T$, becomes invalid for significantly larger $G$. 

Fig.1 presents $\delta\phi$ and $\delta\rho$.  Note that $\delta\phi$ is
monotonic, whereas $\delta\rho$ changes sign, so that the system can be overall
neutral.  Fig.2 presents the number densities $n$ and $p$ associated with
$\delta\rho$ of Fig.1.  The curves sketched (but not computed) in 
Ref.\onlinecite{chaz} (Fig.V.3) agree qualitatively with Figs.1 and 2.   

The individual charge densities $\rho$ due to the diffusion-recombination and
screening modes are presented in Fig.3.  Note that $\delta n_S/\delta p_S=-1$
and $\delta n_D/\delta p_D=1.306$ for the present parameters. (Increasing both
mobilities by a factor of ten would give $\delta n_D/\delta p_D=1.022$.) Hence
quasineutrality is not a good approximation for either mode. The charge density
for the screening mode is positive and concentrated near the surface, whereas
the charge density for the diffusion-recombination mode is negative and more
extended. Note that numerical solutions of the macroscopic transport equations
for semiconductors normally do not assume quasineutrality, so they retain the
full physics.\cite{kormay}  However, they do not permit extensive physical
interpretation.

Fig.4 presents the bulk charge density $\delta\rho$ for $\mu_n$ and $\mu_p$ as
above ($\mu_p/\mu_n=0.2$), and for three additional values of $\mu_p/\mu_n$. 
$\delta\rho$ and $\phi_{Dem}$ increase with the difference in the electron and
hole mobilities.

\begin{figure}
\centerline{\psfig{file=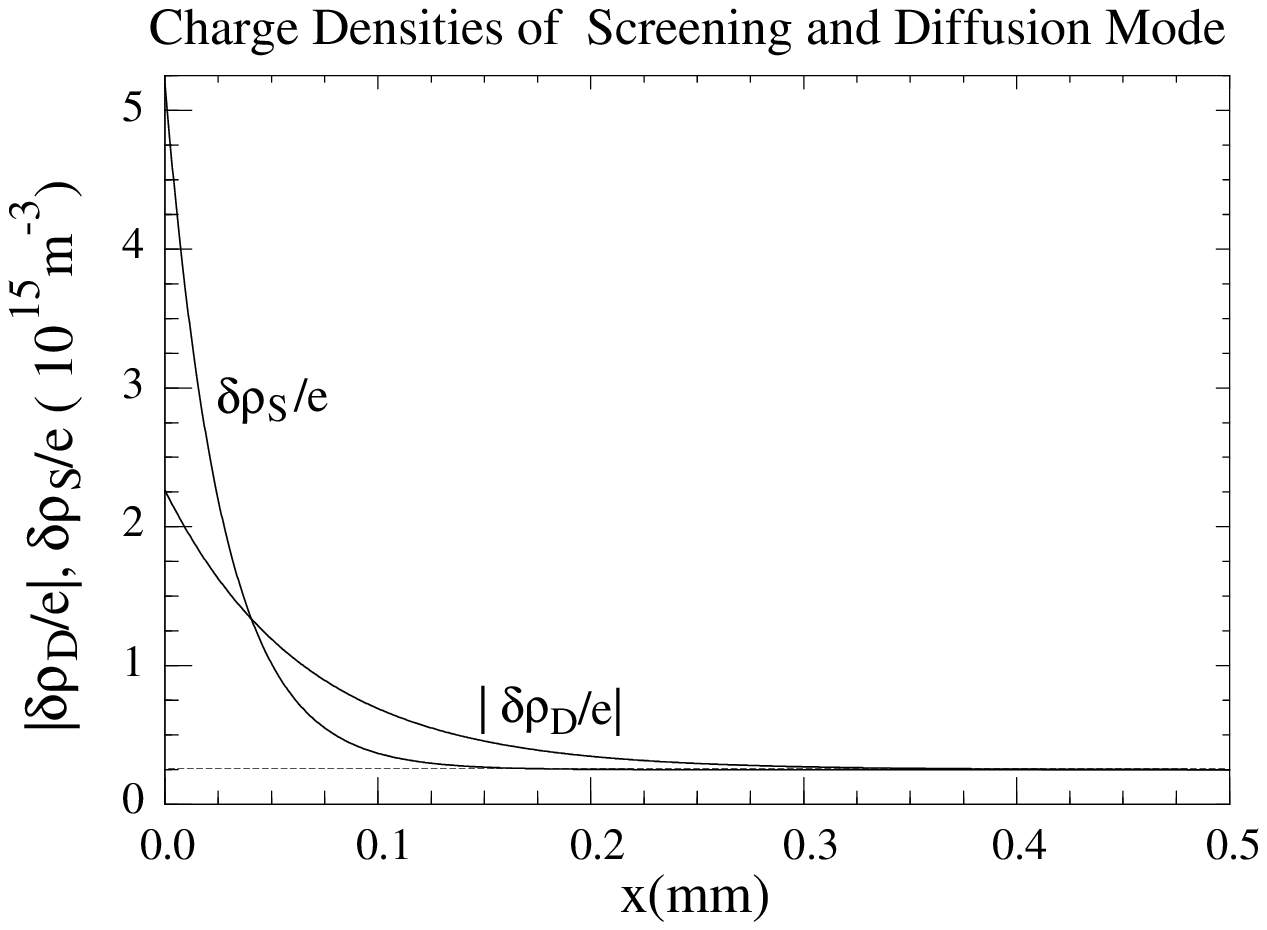,width=1\linewidth}}
\caption{Charge densities for the screening and diffusion modes, from (17) and 
(24).} 
\label{Fig.3}
\end{figure}

\begin{figure}
\centerline{\psfig{file=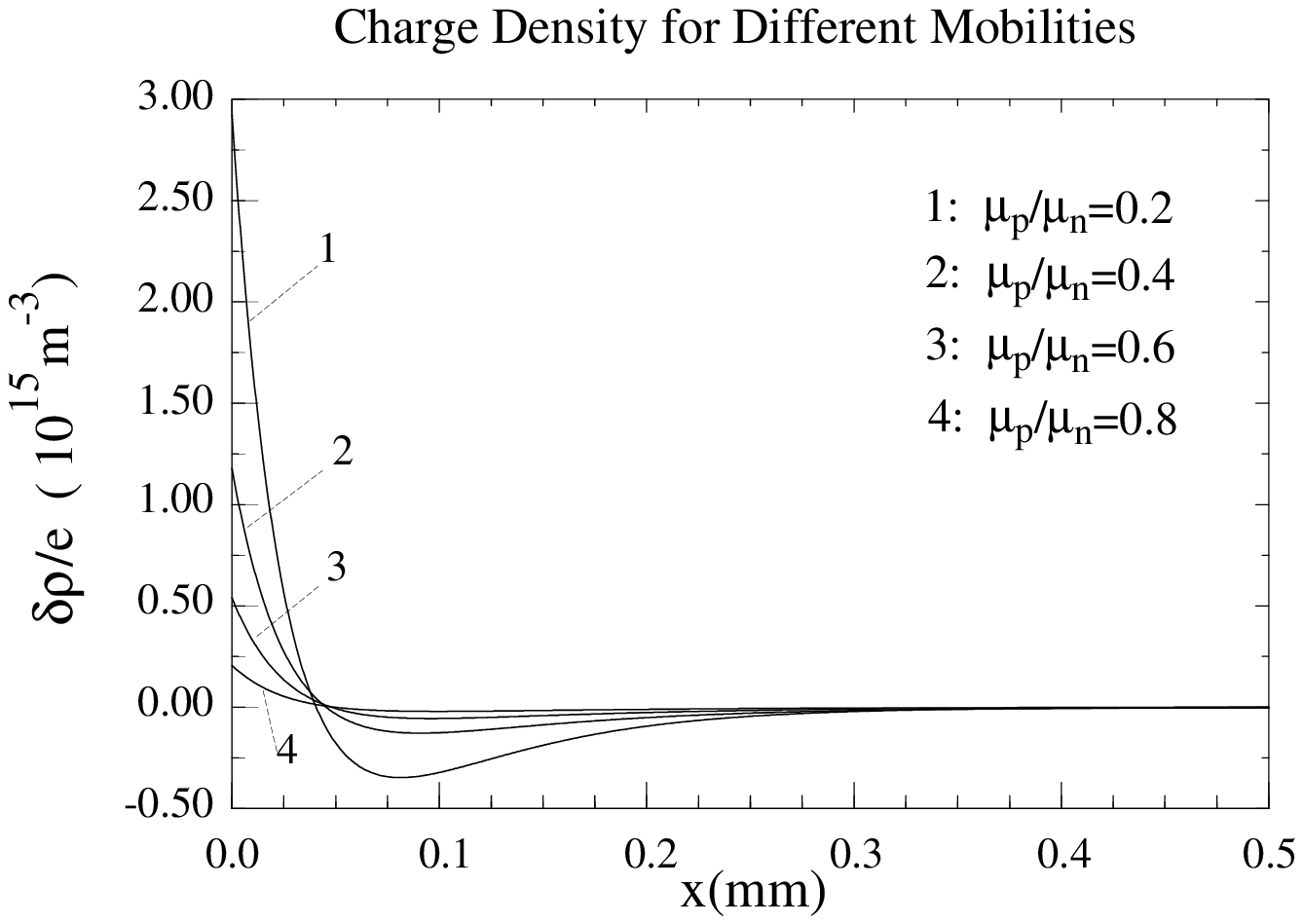,width=1\linewidth}}
\caption{Charge density $\delta\rho$ from (27) for different hole mobilities.} 
\label{Fig.4}
\end{figure}


Surface solutions also are relevant to current flow across interfaces.  Already
in the nineteenth century, Maxwell noted that for current flow between two
single-carrier materials charge builds up at the interface; however, without a
more microscopic theory it is not possible to determine how much charge is
associated with each material.  A similar effect must occur at a
semiconductor-metal interface, even without the complexities associated with
modification of the Schottky barrier. Surface solutions likely are present in
recent numerical studies showing that spin injection in a semiconductor (e.g.,
by preferential absorption of polarized light) persists across a $p-n$
junction.\cite{zutic}  Such a system, with up and down electrons and with
unpolarized holes, should have three surface modes: one screening mode and two
diffusion-recombination modes.  The rapid falloff of spin polarization at one
end of the system is evidence for surface solutions induced by what the authors
of Ref.~\onlinecite{zutic} consider to be an artificial boundary condition.

The present multi-carrier system, with a finite diffusion-recombination length
$L$, is non-neutral only over a finite distance from the disturbance.  On the
other hand, a multi-carrier system with no recombination (and thus
$L\rightarrow\infty$) can be non-neutral in the bulk, far from a
disturbance.\cite{sas}

We would like to thank A. Zangwill for a critical reading of the manuscript. WMS
would like to acknowledge the Department of Energy for its support from DOE
Grant No. DE-FG03-96ER45598.

\end{document}